\documentclass[twoside]{article}

\usepackage[a4paper]{geometry}
\usepackage[utf8]{inputenc}
\usepackage[T1]{fontenc} 
\usepackage{RR}
\usepackage{hyperref}
\usepackage{amsmath,amssymb,amsfonts}
\usepackage{enumitem}
\usepackage{float} 

\newcommand{\conc}{\ensuremath{\,\|\,}}
\newcommand{\Or}{\vee}
\newcommand{\card}{\mbox{card\,}}
\newcommand{\I}[1]{\mbox{\it #1}}

\definecolor{grey}{cmyk}{0,0,0,0.7}

\hypersetup{
pdfborder= 0 0 0,
colorlinks=true,
anchorcolor=grey,
citecolor=grey,
filecolor=grey,
linkcolor=grey,
menucolor=grey,
runcolor=grey,
urlcolor=grey
}

\RRdate{September 2021}

\RRauthor{Pierre Bouvier \and Hubert Garavel}
\authorhead{P. Bouvier \& H. Garavel}
\RRtitle{Le jeu de tests VLSAT-2}
\RRetitle{The VLSAT-2 Benchmark Suite}
\titlehead{The VLSAT-2 Benchmark Suite}
\RRresume{VLSAT-2 (acronyme anglais de ``tr\`es grands probl\`emes de
satisfaisabilit\'e bool\'eenne'') est le second volet d'une suite de tests
destin\'ee aux exp\'erimentations scientifiques et aux comp\'etitions de logiciels
pour la r\'esolution de probl\`emes SAT. VLSAT-2 contient 100~tests
(50~formules satisfaisables et 50~insatisfaisables) de complexit\'e croissante,
fournis en format DIMACS CNF sous une licence Creative Commons permissive.
25\% de ces tests ont \'et\'e utilis\'es lors des \'editions~2020 et~2021
de la comp\'etition internationale sur la r\'esolution SAT.
}
\RRabstract{
This report presents VLSAT-2 (an acronym for ``Very Large Boolean SATisfiability
problems''), the second part of a benchmark suite to be used in scientific
experiments and software competitions addressing SAT-solving issues.
VLSAT-2 contains 100~benchmarks (50~satisfiable and 50~unsatisfiable formulas)
of increasing complexity, proposed in DIMACS CNF format under a permissive 
Creative Commons license. 25\% of these benchmarks have been used during the
2020~and~2021 editions of the International SAT Competition.
}
\RRmotcle{DIMACS CNF, ensemble de donn\'ees, formule SAT, Nested-Unit Petri Net, NUPN, probl\`eme SAT, r\'eseau de Petri, satisfaisabilit\'e bool\'eenne, suite de tests}
\RRkeyword{benchmark suite, Boolean satisfiability problem, data set, DIMACS CNF, Nested-Unit Petri Net, NUPN, Petri Net, SAT formula, SAT solving}
\RRprojet{CONVECS}
\URRhoneAlpes

\begin{document}
\RTNo{0514} 
\makeRT

\section{Benchmark Description}

VLSAT-2\footnote{\url{https://cadp.inria.fr/resources/vlsat/2.html}}
is a collection of 100 SAT formulas. Many of these formulas are difficult to 
handle by current SAT solvers. One half of these formulas is satisfiable, 
while the other half is not.

Each formula is provided as a separate file, expressed in Conjunctive Normal
Form and encoded in the DIMACS CNF
format\footnote{\url{http://www.satcompetition.org/2009/format-benchmarks2009.html}}.
Each file is then compressed using bzip2 to save disk space and allow
faster downloads. The 100 formulas require 5.2~gigabytes of disk space
and 1.4~gigabytes when compressed using bzip2.

The VLSAT-2 benchmarks are licensed under the CC-BY Creative Commons 
Attribution 4.0 International License\footnote{License terms available from 
\url{http://creativecommons.org/licenses/by/4.0}}.

25\% of the VLSAT-2 benchmarks have been selected by the organizers of recent SAT 
Competitions: 7~satisfiable and 7~unsatisfiable formulas have been chosen for 
the SAT~Competition~2020, and 5~satisfiable and 8~unsatisfiable formulas
have been chosen for the SAT~Competition~2021~\cite{Bouvier-Garavel-21-b}.

\section{Scientific Context}
\label{CONTEXT}

Interesting Boolean formulas can be generated as a by-product of our recent
work \cite{Bouvier-Garavel-PonceDeLeon-20} on the decomposition of Petri nets 
into networks of automata, a problem that has been around since the early 
70s. Concretely, we developed a tool chain that takes as input a
Petri net (which must be ordinary, safe, and hopefully not too large)
and produces as output a network of automata that execute concurrently
and synchronize using shared transitions. Precisely, this network is
expressed as a {\em Nested-Unit Petri Net\/} (NUPN)~\cite{Garavel-19},
i.e., an extension of a Petri net, in which places are grouped into sets
(called {\em units\/}) that denote sequential components. A NUPN provides
a proper structuring of its underlying Petri net, and enables formal 
verification tools to be more efficient in terms of memory and CPU time.
Hence, the NUPN concept has been implemented in many tools
and adopted by software competitions, such as the Model Checking 
Contest\footnote{\url{https://mcc.lip6.fr}}
\cite{Kordon-Garavel-et-al-16,Kordon-Garavel-et-al-18} and the Rigorous
Examination of Reactive Systems challenge\footnote{\url{http://rers-challenge.org}}
\cite{Jasper-Fecke-Steffen-et-al-17,Steffen-Jasper-Meijer-vandePol-17,Jasper-Mues-Murtovi-et-al-19}.
Each NUPN generated by our tool chain is {\em flat}, meaning that its units
are not recursively nested in each other, and {\em unit-safe}, meaning that
each unit has at most one execution token at a time.

Our tool chain works by reformulating concurrency constraints on Petri nets
as logical problems, which can be later solved using third-party software,
such as SAT solvers, SMT solvers, and tools for graph coloring and finding 
maximum cliques \cite{Bouvier-Garavel-PonceDeLeon-20}. We applied our approach
to a large collection of more than 12,000 Petri nets from multiple sources,
many of which are related to industrial problems, such as communication protocols,
distributed systems, and hardware circuits. We thus generated a huge collection
of Boolean formulas, from which we carefully selected a subset of formulas
matching the requirements of the SAT Competition.

\section{Structure of Formulas}
\label{STRUCTURE}

Each of our formulas was produced for a particular Petri net. A formula 
depends on three factors: 
\begin{itemize}[itemsep=-0.4ex]
\item the set $P$ of the places of the Petri net;
\item a {\em concurrency relation\/} $\conc$ defined over $P$, such that
$p \conc p'$ iff both places $p$ and $p'$ may simultaneously have 
an execution token; and
\item a chosen number $n$ of units. 
\end{itemize}

A formula 
expresses whether there exists a partition of $P$ into $n$ subsets $P_i$ 
($1 \leq i \leq n$) such that, for each $i$, and for any two places 
$p$ and $p'$ of $P_i$, $p \neq p' \!\!\implies\!\! \neg \, (p \conc p')$.
A model of this formula is thus an allocation of places into $n$ units,
i.e., a valid decomposition of the Petri net. 
This can also be seen as an instance of the graph coloring problem, in which 
$n$ colors are to be used for the graph with vertices defined by the places
of $P$ and edges defined by the concurrency relation.
A formula is only satisfiable if the value of $n$ is large enough (namely,
greater than or equal to the chromatic number of the graph), so that at least 
one decomposition exists.

More precisely, each formula was generated as follows.
For each place $p$ and each unit $u$, we created a propositional variable
$x_{pu}$ that is true iff place $p$ belongs to unit $u$. We then added
constraints over these variables:
\begin{itemize}
\item For each unit $u$ and each two places $p$ and $p'$ such that $p \conc p'$
and $\#p < \#p'$, where $\#p$ is a bijection from places names to the interval
$[1, \card(P)]$, we added the constraint $\neg x_{pu} \Or \neg x_{p'u}$ to
express that two concurrent places cannot be in the same unit.
\item For each place $p$, we could have added the constraint 
$\bigvee_u x_{pu}$ to express that $p$ belongs to at least one unit, but 
this constraint was too loose and allowed $n!$ similar solutions, just by 
permuting unit names. We thus replaced this constraint by a stricter one 
that breaks the symmetry between units: for each place $p$, we added the 
refined constraint
$\bigvee_{1 \le \#u \le \I{\scriptsize min\/} (\#p, n)} x_{pu}$, where
$\#u$ is a bijection from unit names to the interval $[1, n]$.
\end{itemize}

Figure~\ref{GRAPH} illustrates, for a typical Petri net, how the resolution 
time evolves as the chosen number of units increases. If the value of $n$ is
small (resp. large) enough, it is relatively easy to prove that the formula is 
satisfiable (resp. unsatisfiable) because there are not enough (resp. too 
many) colors. The difficulty comes when $n$ gets close to the chromatic number
(which is equal to~13 on Fig.~\ref{GRAPH}), as the number of combinations to 
be examined for proving unsatisfiability explodes, despite the introduction of 
symmetry-breaking constraints, which make unsatisfiable formulas much easier
and satisfiable formulas slightly harder.

\section{Selection of Benchmarks}

Using the approach presented in Sections~\ref{CONTEXT} and~\ref{STRUCTURE}, 
we previously published a test suite, named VLSAT-1~\cite{Bouvier-Garavel-20},
of 100~formulas. However, VLSAT-1 only contains satisfiable formulas, as it 
was designed for the Model Counting Competition, which seeks formulas 
accepting a large number of models.
For the SAT Competition, we therefore undertook the production of a different 
collection containing both satisfiable and unsatisfiable formulas, depending
on the number of units chosen for a given Petri net. 

We selected
50~satisfiable and 50~unsatisfiable formulas, carefully chosen among a large
collection of more than 132,000 formulas generated by our tool chain. 
We used five SAT solvers (namely, CaDiCal 1.3.0, Kissat 1.0.3, MathSAT 5.6.5,
MiniSAT 2.2.0, and Z3 4.8.9) to reject all formulas that can be solved in 
less than one minute of CPU time
by at least one of these solvers, and that can be solved
within two hours by each of these solvers (experiments done on a
Xeon E5-2630~v4 machine with 256~gigabytes of RAM).
We also tried to provide formulas of increasing complexities,
with a good compromise between the size of a formula and
the time taken by the fastest solver to process this formula.

\begin{figure}[H] 
	\centering
	\includegraphics[width=9.8cm]{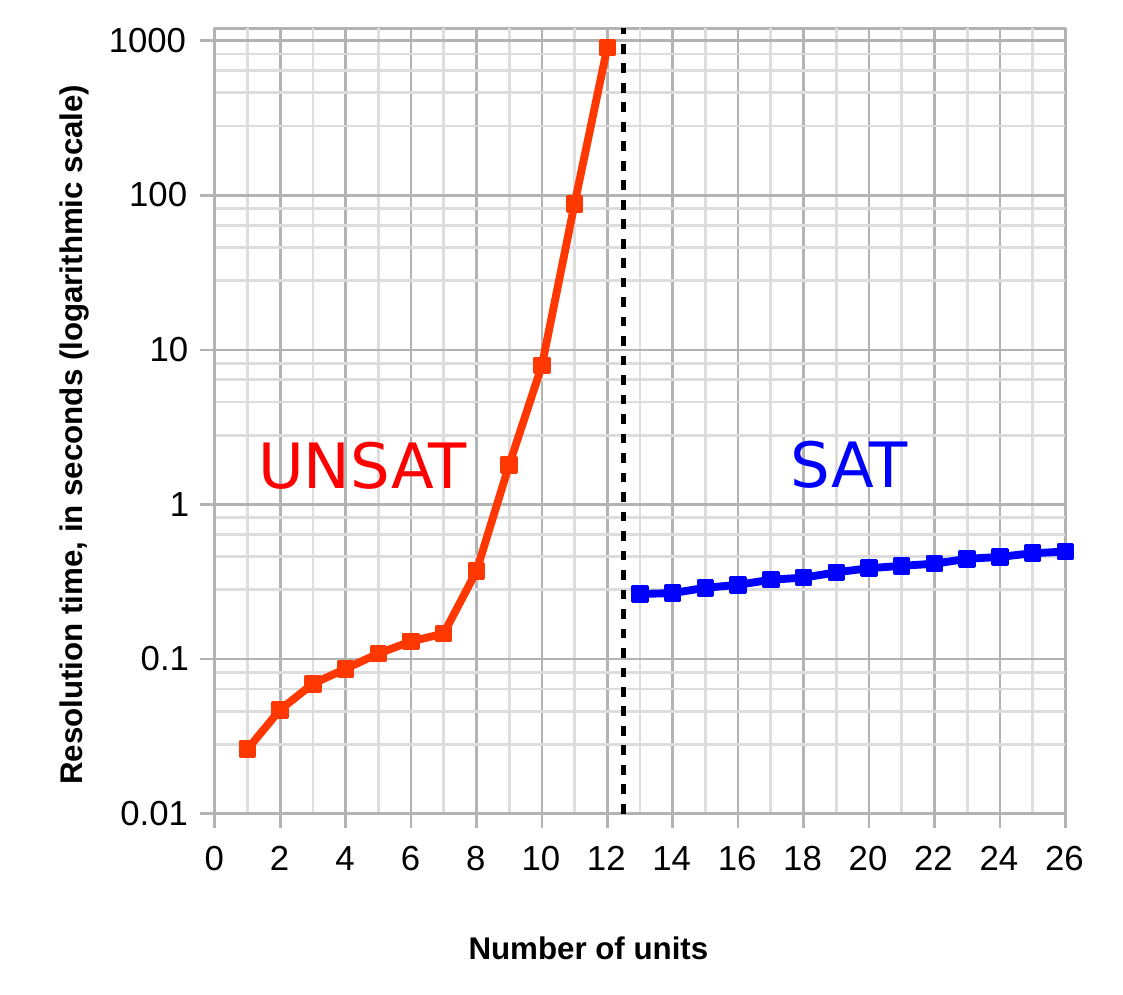}
	\vspace{-0.5cm}
	\caption{Resolution times for a typical NUPN}
	\label{GRAPH}
\end{figure}

\begin{figure}[H] 
	\centering
	\includegraphics[width=9.8cm]{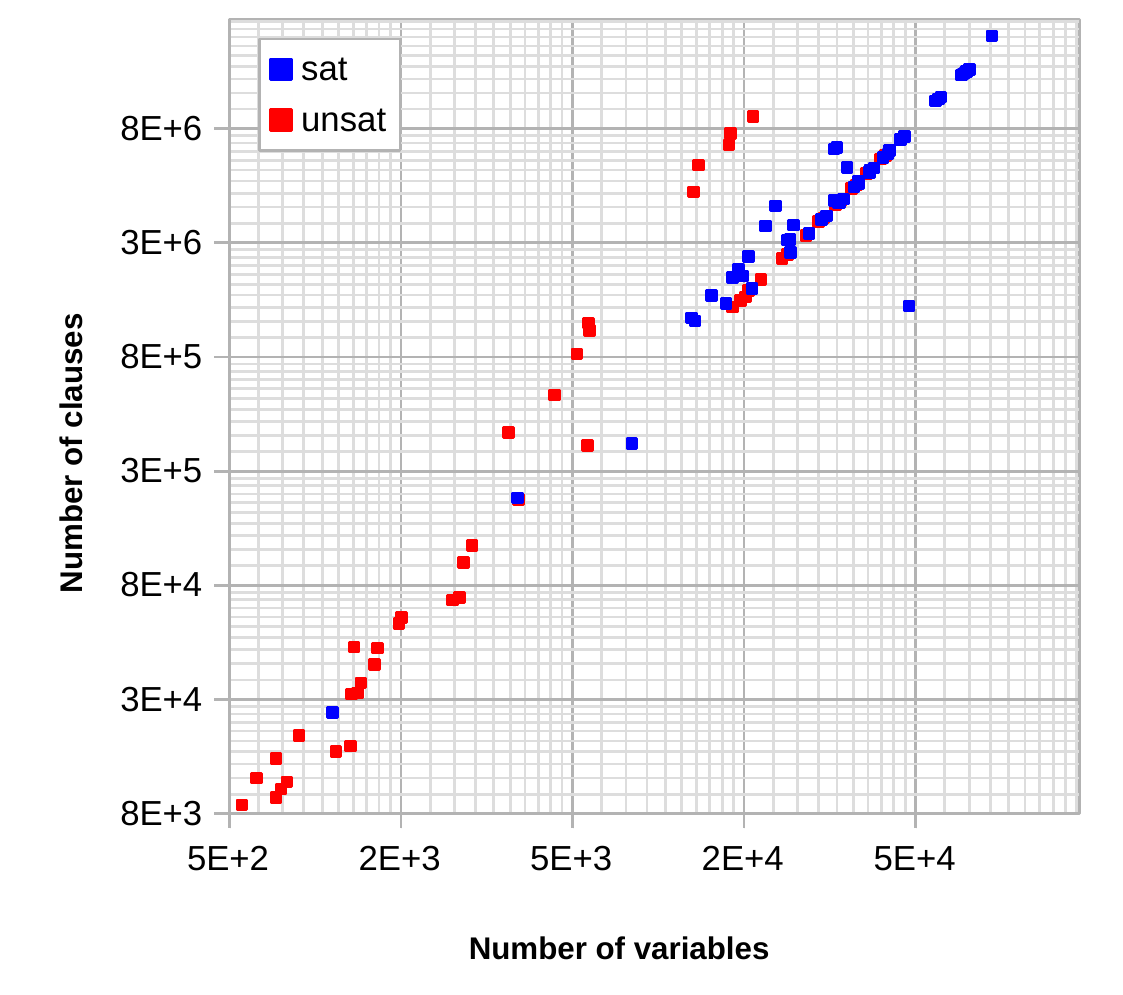}
	\vspace{-0.5cm}
	\caption{Dispersion of the VLSAT-2 benchmarks}
	\label{DISPERSION}
\end{figure}

\begin{table}[H] 
\centering
\setlength{\tabcolsep}{5pt}
\small
\begin{tabular}{|r|r|l|c|} \hline
{\em variables} & \multicolumn{1}{|c|}{\em clauses} & \multicolumn{1}{|c|}{\em type} & {\em difficulty} \\ \hline \hline
544 	& 8738		& UNSAT		& 3 \\
600 	& 11,440	& UNSAT		& 3 \\
684 	& 9417		& UNSAT		& 3 \\
684 	& 13,953	& UNSAT		& 2 \\
708 	& 10,259	& UNSAT		& 3 \\
736 	& 11,022	& UNSAT		& 2 \\
798 	& 17,543	& UNSAT		& 3 \\
1000 	& 22,250	& SAT  		& 4 \\
1022 	& 14,955	& UNSAT		& 3 \\
1125 	& 15,795	& UNSAT		& 3 \\ \hline
1134 	& 26,703	& UNSAT*	& 4 \\
1155 	& 42,917	& UNSAT*	& 4 \\
1183 	& 26,975	& UNSAT		& 3 \\
1209 	& 29,889	& UNSAT		& 3 \\
1326 	& 35,956	& UNSAT		& 3 \\
1352 	& 42,432	& UNSAT		& 3 \\
1560 	& 54,564	& UNSAT		& 3 \\
1586 	& 57,751	& UNSAT		& 3 \\
2231 	& 68,844	& UNSAT		& 2 \\
2340 	& 70,758	& UNSAT		& 2 \\ \hline
2403 	& 100,259	& UNSAT		& 2 \\
2548 	& 119,204	& UNSAT		& 3 \\
3252 	& 372,331	& UNSAT		& 3 \\
3456 	& 192,912	& SAT  		& 2 \\
3480 	& 190,496	& UNSAT		& 3 \\
4424 	& 545,056	& UNSAT*	& 3 \\
5152 	& 824,642	& UNSAT*	& 3 \\
5525 	& 327,765	& UNSAT		& 3 \\
5568 	& 1,124,240	& UNSAT		& 3 \\
5600 	& 1,042,700	& UNSAT*	& 3 \\ \hline
452 	& 334,035	& SAT		& 3 \\
11,130 	& 1,186,888	& SAT*		& 0 (148 s) \\
11,280 	& 4,223,777	& UNSAT*	& 3 \\
11,374 	& 1,150,943	& SAT*		& 1 (1802 s) \\
11,664 	& 5,532,624	& UNSAT		& 4 \\
12,690 	& 1,481,522	& SAT		& 2 \\
14,016 	& 1,374,747	& SAT		& 4 \\
14,280 	& 6,781,327	& UNSAT*	& 3 \\
14,424 	& 7,585,190	& UNSAT*	& 4 \\
14,637 	& 1,778,453	& SAT		& 2 \\ \hline
14,640 	& 1,323,246	& UNSAT		& 5 \\
15,249 	& 1,937,993	& SAT		& 2 \\
15,440 	& 1,409,906	& UNSAT$^+$	& 5 \\
15,704 	& 1,804,650	& SAT		& 3 \\
15,960 	& 1,464,039	& UNSAT$^+$	& 5 \\
16,269 	& 2,203,672	& SAT		& 3 \\
16,297 	& 1,562,268	& UNSAT		& 5 \\
16,676 	& 1,598,591	& SAT$^+$	& 2 \\
16,788 	& 9,021,307	& UNSAT		& 3 \\
17,688 	& 1,741,702	& UNSAT		& 5 \\ \hline
\end{tabular}
\hfill
\begin{tabular}{|r|r|l|c|} \hline
{\em variables} & \multicolumn{1}{|c|}{\em clauses} & \multicolumn{1}{|c|}{\em type} & {\em difficulty} \\ \hline \hline
18,250 	& 2,995,915	& SAT		& 2 \\
19,565 	& 3,665,001	& SAT		& 2 \\
20,424 	& 2,157,568	& UNSAT		& 5 \\
21,114 	& 2,240,429	& UNSAT$^+$	& 5 \\
21,190 	& 2,597,791	& SAT		& 3 \\
21,546 	& 2,615,351	& SAT		& 3 \\
21,573 	& 2,289,124	& SAT		& 3 \\
22,032 	& 3,022,731	& SAT		& 2 \\
23,961 	& 2,714,844	& UNSAT		& 5 \\
24,450 	& 2,770,239	& SAT$^+$	& 1 (721 s) \\ \hline
26,104 	& 3,131,630	& UNSAT		& 5 \\
26,606 	& 3,191,844	& SAT  		& 2 \\
26,988 	& 3,251,923	& UNSAT		& 5 \\
27,507 	& 3,314,450	& SAT  		& 3 \\
28,930 	& 6,497,511	& SAT  		& 4 \\
29,040 	& 3,874,024	& SAT  		& 2 \\
29,205 	& 3,712,921	& UNSAT		& 5 \\
29,456 	& 6,615,638	& SAT		& 2 \\
29,736 	& 3,780,419	& SAT*		& 2 \\
29,945 	& 3,796,274	& SAT		& 4 \\ \hline
30,195 	& 3,855,554	& UNSAT$^+$	& 5 \\
30,744 	& 3,925,645	& SAT$^+$	& 5 \\
31,552 	& 5,400,750	& SAT		& 4 \\
32,480 	& 4,362,044	& UNSAT$^+$	& 5 \\
33,040 	& 4,437,242	& SAT		& 4 \\
33,582 	& 4,529,625	& UNSAT		& 5 \\
34,099 	& 4,695,729	& SAT		& 3 \\
34,161 	& 4,607,712	& SAT$^+$	& 4 \\
35,929 	& 5,082,743	& UNSAT$^+$	& 5 \\
36,518 	& 5,166,057	& SAT		& 4 \\ \hline
36,603 	& 5,137,412	& SAT		& 3 \\
36,792 	& 5,273,558	& SAT		& 2 \\
37,758 	& 5,364,539	& SAT$^+$*	& 4 \\
39,552 	& 5,878,762	& UNSAT$^+$	& 5 \\
40,170 	& 5,970,608	& SAT$^+$	& 5 \\
40,896 	& 6,104,639	& UNSAT		& 5 \\
41,535 	& 6,200,014	& SAT		& 4 \\
41,875 	& 6,420,498	& SAT		& 4 \\
45,150 	& 7,165,285	& SAT		& 4 \\
46,440 	& 7,369,989	& SAT		& 2 \\ \hline
47,817 	& 1,337,056	& SAT		& 2 \\
57,038 	& 10,572,502	& SAT$^+$	& 5 \\
58,380 	& 10,775,711	& SAT		& 4 \\
59,204 	& 10,973,962	& SAT		& 4 \\
67,996 	& 13,708,722	& SAT		& 5 \\
68,760 	& 13,862,744	& SAT		& 5 \\
69,524 	& 14,016,766	& SAT		& 5 \\
70,288 	& 14,170,788	& SAT*		& 4 \\
71,816 	& 14,478,832	& SAT		& 3 \\
83,334 	& 20,350,783	& SAT		& 4 \\ \hline
\end{tabular}
\caption{\label{TABLE} List of VLSAT-2 formulas}
\end{table}

The VLSAT-2 formulas are listed in Table~\ref{TABLE}. Those marked with
a plus (resp.~a~star) in the table have been selected by the organizers of
the SAT~Competition~2020 (resp.~2021).
The column ``difficulty'' contains a number from 0 (easy) to 5 (hard)
indicating how many of the five aforementioned SAT solvers failed to solve 
the corresponding formula within two hours. For the easy values 0 and 1,
the average number of seconds taken by the tools that managed
to solve the formula is given.

Figure~\ref{DISPERSION} shows the dispersion of the VLSAT-2 benchmarks
for both satisfiable and unsatisfiable formulas. In general, and as 
confirmed by Fig.~\ref{GRAPH}, satisfiable formulas need to be much larger 
(in the number of variables and clauses) than unsatisfiable ones to reach 
the same level of difficulty.

\subsubsection*{Acknowledgements}
The experiments presented in this paper were carried out using the 
{\sc Grid'5000}\footnote{\url{https://www.grid5000.fr}} testbed, supported 
by a scientific interest group hosted by {\sc Inria} and including {\sc Cnrs},
{\sc Renater} and several Universities as well as other organizations. 

\begin{small}

\end{small}

\end{document}